\documentstyle[twocolumn,aps,psfig]{revtex}
\begin{document}
\draft
\title{The kinesin walk: a dynamic model with elastically coupled heads}
\author{Imre Der\'enyi$^*$ and Tam\'as Vicsek$^\dagger$}
\address{Department of Atomic Physics,
E\"otv\"os University, Budapest, Puskin u 5-7, 1088 Hungary \\
  {\tt $^*$derenyi@hercules.elte.hu} {\rm ~~and~~}
  {\tt $^\dagger$h845vic@ella.hu} }

\address{~\\ \large \rm Proceedings of the National Academy of Sciences USA
{\bf 93}, 6775-6779 (1996)}

\maketitle

{\noindent \bf ABSTRACT~~~
Recently individual two-headed kinesin molecules have been studied in
{\it in vitro} motility assays revealing a number of their peculiar
transport properties.  In this paper we propose a simple and robust
model for the kinesin stepping process with elastically coupled Brownian
heads showing all of these properties.  The analytic and numerical
treatment of our model results in a very good fit to the experimental
data and practically has no free parameters.  Changing the values of the
parameters in the restricted range allowed by the related experimental
estimates has almost no effect on the shape of the curves and results
mainly in a variation of the zero load velocity which can be directly
fitted to the measured data.  In addition, the model is consistent with
the measured pathway of the kinesin ATPase.
}

\bigskip
\hrule
\bigskip

\begin{figure}
\centerline{\psfig{figure=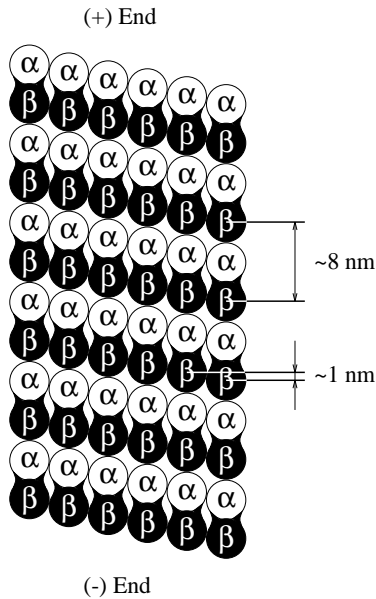,height=3.2in}}
\caption{
The structure of the B-type helical surface lattice of the microtubule. The
kinesin heads can bind only to the $\beta$-tubulin monomers.
}\label{ftub}\end{figure}

\noindent
Kinesin is a motor protein converting the energy of ATP hydrolysis into
mechanical work while moving large distances along microtubule filaments and
transporting organelles and vesicles inside the cytoplasm of eukaryotic cells
\cite{dlb}. The wall of a microtubule is made up of tubulin heterodimers
arranged in 13 longitudinal rows called protofilaments (Fig.\ \ref{ftub}).
A tubulin heterodimer is 8~nm long and consists of two globular
proteins about 4~nm in diameter: $\alpha$- and $\beta$-tubulin. The dimers
bind head-to-tail giving the polarity to the protofilaments. The microtubule
has a helical surface lattice which can have two possible configurations: the
adjacent protofilaments can be shifted in the longitudinal direction by
about either 5~nm (A-type lattice) or 1~nm (B-type lattice). Electron
micrograph measurements of microtubules decorated with kinesin head fragments
\cite{song,hoe,kikk,hir} indicate the predominance of the B-type lattice
and show that kinesin heads can bind only to the $\beta$-tubulin
(meanwhile weakly interacting with the $\alpha$-tubulin also).
Native kinesin is a dimeric molecule with two globular
($\sim$9$\times$3$\times3$~nm) heads. Each one has an ATP and a tubulin
binding site.

Recent experimental studies in {\it in vitro} motility assays have revealed
the following properties of kinesin movement:
\begin{itemize}
\item kinesin moves unidirectionally parallel to the protofilaments towards
the ``$(+)$'' end of the microtubule \cite{Vale,ray};
\item under an increasing load the speed of the kinesin decreases almost
linearly \cite{hunt,sv1} (see Fig.\ \ref{ff_v});
\item under its stall force (about 5~pN) kinesin still consumes ATP at
an elevated rate \cite{sv1};
\item in the absence of ATP (in rigor state) kinesin binds to the microtubule
very strongly: it supports forces in excess of 10~pN \cite{sv1};
\item the observed step size in the low speed regime (at low ATP or at high
force) is about 8~nm \cite{sv2};
\item some backward slippage was clearly observed \cite{sv1};
\item the displacement variance at saturating ATP and at low load increases
linearly with time, but at (only somewhat more than) half of the rate of
a single Poisson stepping
process with 8~nm step size, implying that one step consists of two sequential
subprocesses with comparable limiting rates \cite{sv3}.
\end{itemize}

Partially motivated by these remarkable experimental advances several
interesting thermal ratchet type models have very recently been
developed for the theoretical interpretation of the related transport
phenomena.  These models
\cite{magn,ajd_pros,pr_aj,ast_bier,doer,mill_dyk,mill,bart} except that
of Ajdari (Ref.\ \cite{ajd}) consider cases in which the internal degree of
freedom of the molecular motors is not taken into account.  The purpose
of the present work is to define and investigate a dynamic model for the
kinesin walk which has an internal degree of freedom and results in a
full agreement with the experimental data for the range of its
parameters allowed by the known estimates for the lower and upper bounds
of these parameters.

There are two basic classes of possible models for the stepping of
kinesin \cite{sv4}.  The first one is the ``Long-Stride'' model in which
the heads are moving along a single protofilament.  The two heads are
displaced from each other by 8~nm.  During the stepping process the back
head passes the bound front head advancing 16~nm.  Then the heads change
their roles and a new step may take place.  However, within this
framework during the long stride the back head has a good chance to bind
to a $\beta$ site belonging to a neighboring track which is closer than
the next $\beta$ site in the same track.  Another problem is that the
low displacement variance observed experimentally cannot be naturally
explained.  This model was studied in detail by Peskin and Oster
\cite{ost}, who by introducing several reaction rate constants found a
reasonable agreement with the experimental results.

The other possibility is the family of ``Two-Step'' models. These models
naturally explain the low displacement variance because of the two sequential
subprocesses. During one cycle one of the two heads takes a 8~nm step first
then the other head steps. Note that the motion of just one head advances the
centroid of the molecule by only 4~nm the
successive steps of the two heads in rapid
pair results in an effective 8~nm step which can be observed in the low speed
regime. This kind of motion gives large stability to the protofilament
tracking. The differences in the ``Two-Step'' models arise from the relative
positions of the heads. If both heads track the same protofilament they are
not able to pass each other, so we can always distinguish the front and the
back head. In this case the distance of the heads alternates between 8 and
16~nm, which seems a bit large for the kinesin molecule with its $9-10$~nm
long heads, but is still possible. (Alternating between 0 and 8~nm is
already impossible because only one head can bind to each $\beta$-tubulin site
at a time). Another quite reasonable situation is when the {\it two heads track
adjacent protofilaments} on a B-type lattice. In this case the heads are
sitting on adjacent $\beta$-tubulins displaced from each other by only about
1~nm in the longitudinal direction, thus either can take the first step
(they can be identical).

\section*{Dynamics of the kinesin walk}

\begin{figure}
\centerline{\psfig{figure=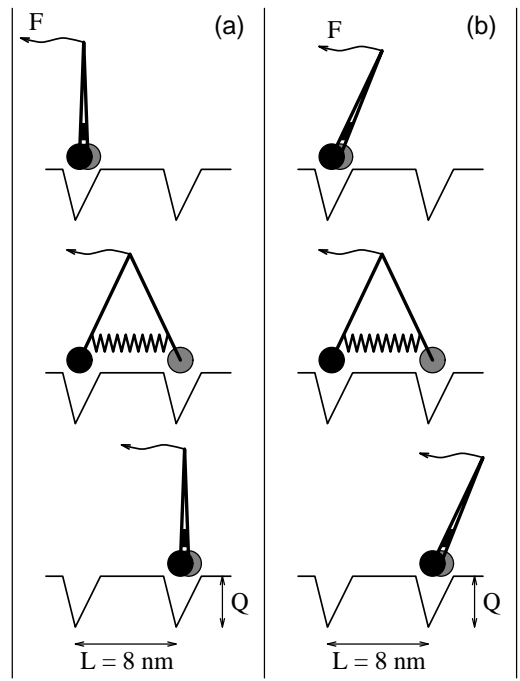,height=3.7in}}
\caption{
Schematic picture of the potential and the subsequent steps (from top to
bottom) of the kinesin molecule in the two limiting cases. In the case (a)
the hinge is always in the centroid of the molecule, and advances
$4-4$~nm during both subprocesses. The two heads share the load force equally.
In the case (b) the relative position of the hinge to the back head is fixed,
thus the whole load force acts on this head. During the first subprocess the
hinge does not move, however, during the second it advances 8~nm. All the
intermediate cases are possible between (a) and (b).
}\label{fwalk}\end{figure}

We present here a one-dimensional kinesin walk model which describes the
whole family of ``Two-Step'' models.  Each of the two Brownian heads can move
along its own one-dimensional periodic potential with period $L=8$~nm in an
overdamped environment.  The two potentials can be shifted relatively to each
other by an arbitrary distance (or even can be the same).  These potentials
represent the interaction with the protofilaments and the periods are the
tubulin heterodimers.  Each period has a deep potential valley corresponding
to the binding site of the $\beta$-tubulin and the other parts of the
potential are flat.  Each valley has an asymmetric ``V'' shape (see Fig.\ 
\ref{fwalk}): the slope in the backward direction (towards the ``$(-)$'' end
of the microtubule) is steeper and $0.5-1$~nm long, while the other slope
(towards the ``$(+)$'' end) is $1.5-2$~nm long, showing the polarity of the
filaments.  These ranges are in the order of the Debye length.

In our model the heads are connected at a hinge, and a {\it spring acts between
them} (Fig.\ \ref{fwalk}). At the beginning of the mechanochemical cycle both
heads are sitting in their valleys waiting for an ATP molecule, and the spring
is unstrained. Any configuration of the model is mathematically equivalent
to a model in which the two potentials are identical, the heads are
sitting in the same valley, and therefore the rest length of the spring is
zero. (Further on we will consider only this version of the model.) After one
of the heads binds an ATP molecule, the hydrolysis of this ATP causes a
conformational change in this head, more precisely, induces the head to take a
8~nm forward step. In the language of the model it means that the rest length
of the spring changes from zero to 8~nm right after the hydrolysis. Then, as
the first rate-limiting subprocess, the strained spring is trying to stretch
pushing the head to the next valley. Reaching its new 8~nm rest length
another conformational change occurs in the head as a consequence of the ADP
release: the rest length of the spring changes back to zero quickly, then, as
the second rate-limiting subprocess, the spring is trying to contract pulling
forward the other head. Completing the contraction a next cycle can start
waiting for a new ATP molecule.
Hirose {\it et al.} \cite{hir} provide evidence that the kinesin$\cdot$ADP
complex has indeed a different conformation near the junction of the heads.
(Similarly, a myosin head also gets over
several conformational changes \cite{raym,ostap}.) An important cooperative
feature of our model is that only one ATP hydrolysis can occur
during each cycle.

This picture is consistent with the scheme of Gilbert {\it et al.} \cite{gilb}
for the pathway of the kinesin ATPase. The first rate-limiting subprocess was
observed as the dissociation of the head from the microtubule followed by a
fast rebinding. They did not report on the second one, but it is clear that
this subprocess already does not belong to the chemical part of the cycle,
and induces a slow dissociation and a fast rebinding of the other head.

\begin{figure}
\centerline{\psfig{figure=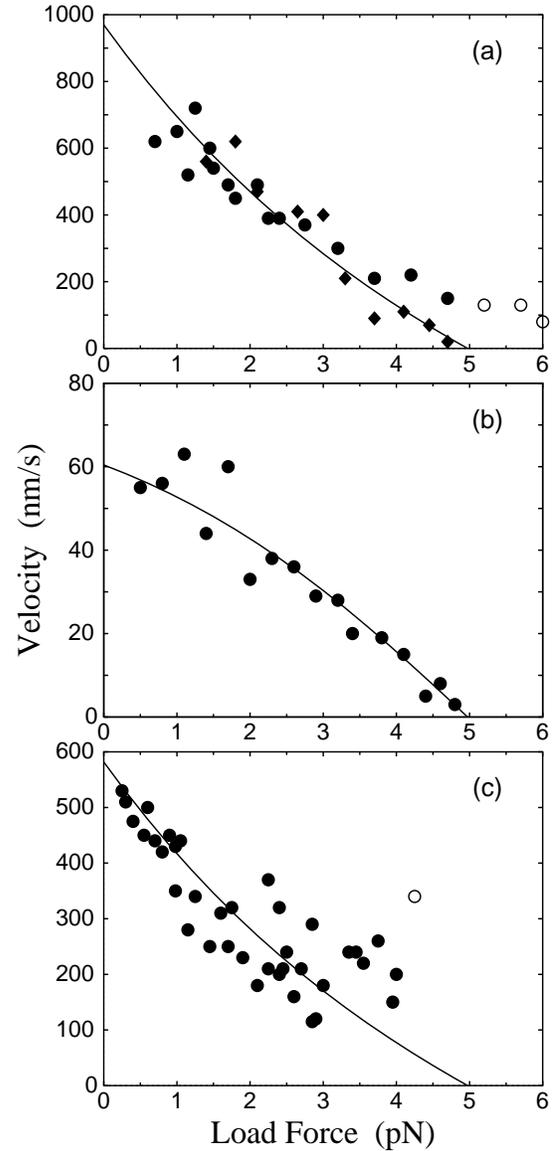,height=6.12in}}
\caption{
The force-velocity curves for an individual kinesin molecule at saturating
($\gg$ 90~$\mu$M) ATP in (a) and (c); and at low ($\sim$10~$\mu$M) ATP
concentration in (b). The experimental data (scattered symbols) are from
Svoboda and Block [9] in (a) and (b) obtained by using optical tweezers; and
from Hunt {\it et al.} [8] in (c) applying viscous load. The open circles
correspond to the simultaneous effect of multiple kinesin motors in the
authors' interpretation [8,9]. The circles and diamonds in (a) mean two
different set of the measured data. To get the best fit (solid lines) we
chose 0.7~nm for the backward length of the potential valleys and 1.75~nm for
the forward length; the depth of the valleys was $Q=20kT$. In these plots
the load force was $F/2$ on each kinesin head. In plot (b) we multiplied the
average displacement by the ATP consumption rate $\nu(c_{\mbox{\tiny ATP}})
=10$~1/s for the best fit. In plot (c) we used the same curve as in (a) but
multiplied by a factor 0.6, which can be explained with the larger viscosity.
}\label{ff_v}\end{figure}

We can take the load force into account in a natural way. Since in the
experiments of Svoboda {\it et al.} \cite{sv1,sv2} a large ($\sim$0.5~$\mu$m
in diameter) and therefore slow (compared to the kinesin heads) silica bead
was linked to the hinge of the kinesin molecule by a relatively weak elastic
tether, we can apply a constant (independent of time) force $F$ to the
hinge. We arrive
at the same conclusion by considering the experiments of Hunt {\it et al.}
\cite{hunt}, where
a viscous load acted on the moving microtubule while the long tail of the
kinesin was fixed. But how can we transfer the force to the heads
when writing down the equations of their motion? There are two possible
limiting cases: if the hinge is always in the centroid of the molecule (Fig.\ 
\ref{fwalk}a) the two heads share the load force equally, {\it i.e.}, $F/2$
force acts on both heads; if the head which wants to step is always free from
the load force (as in Fig.\ \ref{fwalk}b) the whole force $F$ acts on the
actual back head. Due to the robustness of the model both (and therefore all
the intermediate) cases yield practically the same force-velocity curves
(Fig.\ \ref{ff_v}).

The motion of the heads are described by the Langevin equations
%
\begin{eqnarray}
&&\Gamma\dot x_1 = -\partial_{x}V(x_1)-F_1^{\rm load}
 +K\cdot(x_2-x_1-l(t))+\xi_1(t) \, ,
\nonumber\\
&&\Gamma\dot x_2 = -\partial_{x}V(x_2)-F_2^{\rm load}
 +K\cdot(x_1-x_2-l(t))+\xi_2(t) \, .
\nonumber\\ \label{lang}
\end{eqnarray}
where $x_1$ and $x_2$ denote the positions of the heads; $\Gamma$ is
the frictional drag coefficient; $V(x)$ is the periodic potential; $l(t)$ is
the rest length (which alternates between 0 and 8~nm due to the conformational
changes) and $K$ is the stiffness of the spring; $\xi_1(t)$, $\xi_2(t)$ are
Gaussian white noises with the autocorrelation function
$\bigl<\xi_j(t)\xi_i(t')\bigr>=2kT\Gamma\delta_{j,i}\delta(t-t')$
for $i,j=1,2$.

\section*{Results}

In order to obtain results we can compare with the experiments first the values
of the input parameters have to be specified.
The drag coefficient $\Gamma$ for a single head can be calculated from
the Stokes formula yielding $\Gamma\approx 6\times 10^{-11}$~kg/s.

The free energy which can be gained from the ATP hydrolysis is about $25kT$ or
$100\times10^{-21}$~J. During a stepping cycle the two conformational changes
consume the free energy of the ATP, while the rest length of the spring
changes by 8~nm in both cases. This means that $1/2\cdot 25kT \approx
(K/2)(8$~nm$)^2$, from which we get $K\approx1.5$~pN/nm.

A lower limit for the depth $Q$ of the potential valleys can be determined
from the fact that the kinesin in rigor state supports forces in excess of
10~pN. If we try to pull out the two-headed kinesin molecule (with drag
coefficient $2\Gamma$) from a $2Q$ deep potential valley with a force $F$, a
low limit for the escape rate is

\begin{displaymath}
{F^2\over kT2\Gamma} e^{-2Q/kT}
\end{displaymath}

\noindent as follows from a more general expression (\ref{esc}) to be
discussed later. Assuming that the escape rate is much larger than
$10^{-2}$~1/s we get $13kT$ as a lower limit of $Q$. In reality the two heads
cannot be handled as one larger head, because they are not fixed to each
other too rigidly, therefore, this is a very low limit. $Q\approx20kT$
is expected to be a better estimate.

Note that the temperature $T$ plays an important role in our model due to the
deep potential valley the Brownian kinesin heads have to escape from.

We can assume that at the beginning of a stepping cycle both heads are
sitting in the same valley of the potential. After the ATP hydrolysis, as the
first conformational change, the strained spring is trying to stretch,
pushing one of the heads to a neighboring valley, which is $L=8$~nm away. If
the load force is small there is a large probability
$p_{0L}^+ = J_{0L}^+/(J_{0L}^+ +J_{0L}^-)$
that the front head jumps to the forward direction due to the asymmetry of
the potential valleys (the forward slope is less steep than the backward),
and there is only a small probability
$p_{0L}^- = J_{0L}^-/(J_{0L}^+ +J_{0L}^-)$
that the other head jumps backward. $J_{0L}^+$ and $J_{0L}^-$ denote the
corresponding jumping rates. Increasing the load the probability of the
forward step decreases while that of the backward step increases. The average
time
$t_{0L}=1/(J_{0L}^+ +J_{0L}^-)$
which is needed for this stretching also slightly increases. Completing this
subprocess the second conformational change occurs: the spring is trying to
contract. Now for low load force the probability
$p_{L0}^+ = J_{L0}^+/(J_{L0}^+ +J_{L0}^-)$
that the backward head jumps forward to the next valley, where the other head
is sitting, is close to 1, while the probability
$p_{L0}^- = J_{L0}^-/(J_{L0}^+ +J_{L0}^-)$
that the forward head jumps backward is very small. Increasing the
load the situation is similar to the previous case.

Thus, under low load force the kinesin molecule steps 8~nm forward during
almost each mechanochemical cycle. But increasing the load the probability
that the molecule remains on the same place or even takes a backward 8~nm step
increases. Reaching the stall load the average displacement of the kinesin
becomes zero.

Since the potential valleys are very deep (compared to the thermal energy
$kT$), the heads spend almost all of their time at the bottom, and they are
able to jump only very seldom. Therefore, we can calculate the above
mentioned jumping rates ($J=J_{0L}^+$, $J_{0L}^-$, $J_{L0}^+$ or
$J_{L0}^-$) from an Arrhenius-like formula. All we need for this is the
effective potential $V^{\rm eff}(x)$ for the jumping head at the bottom of
the valley and at the top of the barrier.
The effective potential consists of five parts:
the $V(x)$ periodic potential for the jumping head;
the potential $F_j^{\rm load}x$ of the external load force;
the spring potential $\bigl<(K/2)(x-x_r-l(t))^2\bigr>_{x_r}$;
the $\bigl<V(x_r)\bigr>_{x_r}$ potential for the remaining head;
and the potential $\bigl<F_r^{\rm load}x_r\bigr>_{x_r}$ of the external load
force for the remaining head.
In this approximation $\bigl<\dots\bigr>_{x_r}$ means thermal averaging
for the position $x_r$ of the remaining head at the bottom of its valley.
Thus, from the corresponding Fokker-Planck equation the jumping rate is

\begin{equation}
J={D\cdot e^{-(V^{\rm eff}_{\rm max}-V^{\rm eff}_{\rm min})/kT} \over
  \int e^{-(V^{\rm eff}_{\phantom{mi}}(x)-V^{\rm eff}_{\rm min})/kT}dx \cdot
  \int e^{-(V^{\rm eff}_{\rm max\phantom{l}}\!\!-V^{\rm eff}_{\phantom{mi}}(x))/kT}dx}
  \; ,
\label{esc}
\end{equation}

\noindent where $D=kT/\Gamma$ denotes the diffusion coefficient;
$V^{\rm eff}_{\rm min}$ and $V^{\rm eff}_{\rm max\phantom{l}}\!\!$ are the
minimal and the maximal values of the effective potential; and the integrals
should be evaluated around the appropriate extremum.
Due to the depth of the potential
valleys if we do not average the position of the remaining head but assume
that it always sits at the bottom, we get similar results.

From the jumping rates one can calculate the average displacement
$d=(p_{0L}^+ p_{L0}^+ -p_{0L}^- p_{L0}^-)L=
   (J_{0L}^+ J_{L0}^+ -J_{0L}^- J_{L0}^-)
     [1/(J_{0L}^+ +J_{0L}^-)][1/(J_{L0}^+ +J_{L0}^-)]L$
during a cycle and the average time
$t=t_{0L}+t_{L0}=1/(J_{0L}^+ +J_{0L}^-)+1/(J_{L0}^+ +J_{L0}^-)$
needed for this. Increasing the load the average
displacement decreases due to the increasing probabilities of the remaining
and the backward slippage; and the average time slightly increases as a
manifestation of the so-called Fenn effect \cite{fenn}.
(At stall load it is about three times larger than without load.)

At saturating ATP concentration the only rate-limiting factor is the stepping
process, therefore dividing the average displacement by the average time gives
the average velocity $v=d/t$ of the kinesin (Fig.\ \ref{ff_v}a and c).

But at low ATP concentration the rate-limiting factor is the diffusion of the
ATP to the kinesin heads, {\it i.e.}, the stepping process is much faster than
getting an ATP. Thus the average velocity is proportional to the average
displacement during one cycle with the prefactor

\begin{displaymath}
\nu (c_{\mbox{\tiny ATP}})={\nu_{\rm sat}\cdot c_{\mbox{\tiny ATP}} \over
  K_{\rm m}+c_{\mbox{\tiny ATP}} }
 \approx{\nu_{\rm sat}\over K_{\rm m}}\cdot c_{\mbox{\tiny ATP}}
 \approx {\rm const} \cdot c_{\mbox{\tiny ATP}}
\; ,
\end{displaymath}

\noindent which is the rate of the ATP consumption. $c_{\mbox{\tiny ATP}}$
denotes here the ATP concentration; $\nu_{\rm sat}$ is the inverse average
time of a cycle; and $K_{\rm m}$ is the mechanochemical Michaelis-Menten
constant. Thus the velocity of the kinesin is $v=\nu (c_{\mbox{\tiny ATP}})d$
(Fig.\ \ref{ff_v}b).

The difference between the shapes of the force-velocity curves at saturating
(Fig.\ \ref{ff_v}a and c) and at low (Fig.\ \ref{ff_v}b) ATP concentration shows
unambiguously the Fenn effect.

Changing the distribution of the load force $F$ between the heads causes the
change of the average time ($t_{0L}$ and $t_{L0}$)
needed for the two subprocesses, but their sum,
{\it i.e.}, the average duration $t$ of the stepping cycle, and the average
displacement $d$ remains practically unchanged. Thus the force-velocity curves
are essentially the same for any distribution of the load force.

We have also solved the Langevin equations (\ref{lang}) by {\it numerical}
integration and our values are in full agreement with the results obtained
from the above analysis.

\section*{Discussion}

Our model describes the stepping process of the kinesin molecule with
elastically coupled heads. We do not use reaction rate constants, but only
strain induced conformational changes and the underlying asymmetric periodic
potential. Therefore, we can give the full description of the dynamics
of the motion.

From Eq.\ (7) of Ref.\ \cite{sv3} it follows that the displacement variance
of a ``Two-Step'' process (in which each cycle consists of two sequential
subprocesses with comparable limiting rates) is $r=1-p_s/2$ times smaller
than that of a single Poisson stepping process taking the same overall time.
The stepping probability $p_s\leq1$ denotes the probability that a cycle
produces a forward step.
In our model at low load force the stepping probability is $p_s=p_{0L}^+
p_{L0}^+ \approx0.8$, in addition, if the ATP concentration is saturated the
two subprocesses are the only rate-limiting factors with the same rates.
This leads to $r\approx0.6$ as compared to $r\approx0.52$ given with an error
in the range of $0.1-0.2$ in Ref.\ \cite{sv3}.
Thus, our model {\it explains the low displacement variance} at
saturating ATP and at low load force. Furthermore, {\it fits the
measured force-velocity
curves} extremely well, and is consistent with the recent experimental
studies of the {\it pathway of the kinesin ATPase}. The parameters of the
model are essentially determined by experiments or theoretical calculations,
therefore, they can only be tuned in a rather restricted range. The shapes of
the force-velocity curves are essentially insensitive to the parameters
within this range, showing the robustness of the model.

If the relative position of the hinge to the backward head is fixed as, {\it
e.g.}, in Fig.\ \ref{fwalk}b then only one power stroke occurs during one
cycle. The main advantage of this
version is that the hinge advances the whole 8~nm distance at once.
(Note, that in this case the first subprocess is not so important, thus it can
be of a different type, for example, similar to the myosin step:
detachment, free swinging, and rebinding to the next side.)
In any other cases (like in Fig.\ \ref{fwalk}a), the two subprocesses
mean two power strokes, and the hinge advances the 8~nm in two parts. In
spite of this, one may observe only 8~nm displacements, as the two
power strokes occur in rapid succession.

%
%
From Eq.\ (\ref{esc}) it follows that the depth
and the degree of asymmetry of the valleys in the periodic potential
we assumed have the strongest effects on
the results. On the other hand, some further details (such as, for
example, a
shallow secondary valley instead of the flat part) have less significant
influence on the force-velocity curves.
%
%
The temperature plays an essential role in the actual mechanism of the
walk.  The two major points where temperature comes into the picture are
the following: i) the potential valley has to be considerably deeper
than $3kT$ to ensure proper binding, ii) on the other hand, from a deep
valley the heads may escape only if the spring and the temperature act
simultaneously since the energy of the spring cannot be larger than that
of the ATP hydrolysis.

To decide which theoretical model describes the kinesin stepping process in
the best way, further experimental studies are needed. One possibility is to
see how the force-velocity curves continue both at higher forces than the
stall force and at negative forces. Another possibility is to measure the
displacement variance at higher loads and at lower ATP concentrations. This
could distinguish among the different cases of our model as well, because for
different distributions of the load force $F$ the average duration of the
subprocesses are also different resulting in different variances.

We are aware that C. Doering, M. Magnasco and G. Oster have considered a
similar approach to the problem of the kinesin walk \cite{DoeMagn}.

\bigskip
The authors are grateful to J. Prost and A. Ajdari for useful
discussions and for their kind hospitality during the visit of I. D. at ESPCI
in Paris. The present research was supported by the Hungarian Research
Foundation Grant No.\ T4439 and No.\ F17246, and the contract
ERB-CHRX-CT92-0063.

\end{document}